\def\beq{\begin{equation}}
\def\seq{\end{equation}}
\def\beqs{\begin{eqnarray}}
\def\seqs{\end{eqnarray}}
\def\lb{\label}
\def\D{\Delta_0}
\def\w{\omega}
\documentstyle[aps,prl,multicol]{revtex}
\begin{document}
\input epsf.sty
\draft 
\title{Anisotropic thermodynamics of $d$-wave
superconductors in the vortex state.}
\author{I. Vekhter$^1$,  P. J. Hirschfeld$^2$, J. P. Carbotte$^3$, and
E. J. Nicol$^1$}
\address{${}^1$Department of Physics, University of Guelph, Guelph, 
	Ontario N1G 2W1, Canada\\
	${}^2$Department of Physics, University of Florida,
	Gainesville, FL 32611, USA\\
	${}^3$Department of Physics and Astronomy, McMaster University,
	Hamilton, Ontario  L8S 4M1,  Canada}
\date{\today}
\maketitle
\begin{abstract}
We show that the density of states and the thermodynamic
properties of a
2D $d$-wave superconductor in the vortex state
with applied magnetic field $\bf H$ in the plane
depend on the angle between $\bf H$ and the
order parameter nodes.  Within a semiclassical
treatment of the extended quasiparticle states,
we obtain fourfold oscillations of the specific
heat, measurement
of which provides a 
simple probe of gap symmetry. The
frequency dependence of the density of states and the
temperature dependence of thermodynamic properties obey different 
power laws for field in the nodal and anti-nodal direction.
The fourfold pattern is changed to twofold when orthorhombicity
is considered.
\end{abstract}

\begin{multicols}{2}

The experimental data accumulated over the last few years have established 
a consensus that
the superconducting state of the hole-doped high-T$_c$ cuprates has
 a predominantly $d$-wave order parameter\cite{kirtley}. 
Such an order parameter 
possesses
lines of nodes, which
results in a gapless excitation spectrum along certain directions
in momentum space. 
An important consequence
 is that the low temperature dependence of 
thermal and transport properties of the 
superconducting cuprates is given by power laws, 
rather than exponential functions with an activation energy
\cite{annett}.

The properties of the vortex state of $d$-wave superconductors also differ 
significantly from those 
of  $s$-wave materials: while for $s$-wave 
the density of states (DOS) and the entropy
are determined at low magnetic fields $H\ll H_{c2}$ by the localized states 
in the vortex cores, in superconductors with lines of nodes they are
dominated by the extended
quasiparticle states, which exist in the bulk along
the nodal directions in momentum space \cite{volovik}.
On the basis of this observation
 K\"ubert and Hirschfeld \cite{kubert} suggested a 
method
of calculating thermal and transport coefficients 
in the vortex state microscopically
by considering only
the contribution of the extended states and accounting 
for the effect of the magnetic field on these states semiclassically,
via a Doppler shift of the quasiparticle 
energy due to the circulating supercurrents. 
For  the field  applied perpendicular to the 
superconducting layers,
the supercurrents can be
approximated by  the circular velocity 
field
around a single vortex, ${\bf v}_s=\hat\beta/2mr$, 
where $r$ is the distance from the center of the vortex, and we have
set $\hbar=1$. Here
$\hat\beta$  is a unit vector along the current and we use
$\beta$ as the vortex winding angle. This expression 
is valid outside the vortex core and up to a cutoff of order
$\min\{R,\lambda\}$, where $2R=2a^{-1}\sqrt{\Phi_0/\pi H}$ 
is the intervortex spacing, $\Phi_0$ is the flux quantum, $a$
is a geometric constant, and
$\lambda$ is the penetration depth. Under these assumptions the
energy of a quasiparticle with momentum {\bf k} is shifted by
$\delta\omega_{\bf k}({\bf r})={\bf v}_s\cdot {\bf k}$,
and the calculated
physical quantities depend on position and have
to be averaged over a unit cell of the vortex lattice.
The results obtained within this framework 
\cite{kubert,kubert2,vekhter,barash} 
describe
recent experiments well \cite{moler,chiao,aubin,sonier,phillips,junod}, 
suggesting that,  
 while the effects left outside of this approach
are important for a quantitative analysis,
the method proposed in Ref.\cite{kubert} is qualitatively correct
and can be used to analyze the properties of the vortex state
of unconventional superconductors.

In this paper we generalize the approach of Ref.\cite{kubert}
to consider the experimental arrangement
 with the magnetic field $H$ in the CuO$_2$ plane, 
and calculate the density of states. 
We find that it exhibits 
fourfold oscillations as a 
function of the direction of the applied field, and that
its energy dependence is drastically
different for the field directed along the node and  anti-node.

Since the c-axis coherence length, $\xi_c$, is shorter than the interlayer
distance, $s$,  
the structure of the vortex state for $\bf H$ in the plane differs from that
with ${\bf H}\|c$, and is commonly modeled by treating the incoherent c-axis
transport as Josephson tunneling 
between the layers.\cite{clem1,bulaevskii,clem2}
The conclusion of Ref.\cite{volovik} that the extended states dominate the
thermodynamic properties of the vortex state 
with lines of nodes in the gap remains valid
for any orientation of the magnetic field. The superfluid velocity,
${\bf v}_s$, away from the core is governed solely by the $2\pi$-winding
of the phase of the order parameter around each vortex, and, at distances large
compared to the size of the core, is virtually identical  to that of an 
Abrikosov vortex.\cite{clem1,bulaevskii}
Since the core size is larger than $\xi_c$, the velocity field
 can be approximated
by  the superflow around a single vortex only for fields
$H\ll H_0=\Phi_0/\gamma s^2$, where $\gamma$ is the anisotropy ratio;
above that field the cores begin to overlap.\cite{bulaevskii}
While for extremely anisotropic cuprates, such as 
Bi$_2$Sr$_2$CaCu$_2$O$_{8}$ family, 
the crossover field is of order of a few Tesla,
for less anisotropic materials, such as YBa$_2$Cu$_3$O$_{7-\delta}$,
  $H_0\ge 50$ Tesla, and for all 
experimentally relevant fields the individual vortices can be treated as
Abrikosov vortices in the calculation of bulk properties.
Furthermore, in the regime $H\ll H_0$ the intervortex distance and
the structure of the vortex state are asymptotically close to those of
an Abrikosov vortex lattice\cite{clem2}, suggesting the the approach of 
Ref.\cite{kubert} can be directly applied to the geometry with the field in the
plane. Finally, the c-axis transport remains incoherent at low temperatures
\cite{bonn}, and therefore only the energies of the quasiparticles 
with momenta
in the 
plane are relevant to the thermodynamic properties and
should be Doppler shifted.

We now follow the approach of
Ref.\cite{kubert} in
neglecting the contribution of the core states and 
assuming a spatially uniform order parameter $\Delta_{\bf k}$ over a 
cylindrical Fermi surface. In most of this work we consider a pure d-wave
angular dependence of the gap, 
$\Delta_{\bf k}\equiv\Delta_0  f(\phi)=\Delta_0\cos2\phi$.
We consider a magnetic field ${\bf H}$ applied in the a-b plane, at an angle
$\alpha$ to the x-axis, and account for its effect on the extended 
quasiparticle states by the Doppler energy shift
$\delta\omega_{\bf k}({\bf r})={\bf v}_s\cdot {\bf k}$. 
The superfluid velocity is approximated by the 
flow field of an isolated vortex,
which is elliptical due to the anisotropy 
of the penetration depth. We can however rescale the c-axis 
 to make both ${\bf v}_s$ and
the intervortex spacing $2R$
isotropic, in the London theory this rescaling is
$z^\prime=z(\lambda_{ab}/\lambda_c)$,
and 
since
the Fermi surface is two-dimensional there is no 
associated rescaling of momentum. Then, 
approximating the unit cell of the vortex lattice by a circle of radius $R$,
we obtain
\begin{equation}
\label{doppler}
\delta\omega_{\bf k}({\bf r})={E_H\over \rho}\sin\beta\sin(\phi-\alpha),
\end{equation}
where  
$\rho=r/R$ and $E_H$ is the energy scale associated with the Doppler shift
\begin{equation}
\label{eh}
E_H={a\over 2} v^*\sqrt{\pi H\over \Phi_0}.
\end{equation}
Here $a$ is a geometric constant of order unity, and
in the London theory the rescaled Fermi velocity
$v^*=v_f\sqrt{\lambda_{ab}/\lambda_c}$, where $v_f$ is 
the Fermi velocity in the plane. In a more 
general approach $v^*$ should be treated as  a parameter
related both to $v_f$ and the anisotropy of the vortex lattice.

The main difference between geometric arrangements
with the field applied in the plane and that
applied along the c-axis is clearly seen from Eq.(\ref{doppler}).
For the 
field applied perpendicular to the layers the momentum and real
space degrees of freedom decouple\cite{kubert}, 
and the average Doppler shift 
is the same at all points on the Fermi surface. 
In contrast, for the field in the plane the average Doppler shift
becomes dependent on the position at the Fermi surface, and is
given by $E_H\sin(\phi-\alpha)$. Since
quasiparticles contribute to the density of states when their Doppler shifted
energy exceeds the local energy gap, 
an immediate conclusion is that the density of states depends sensitively
 on the angle between the applied field and the direction  of
 nodes of the order parameter.

To analyze this dependence quantitatively
we employ the single-particle Green's function which is obtained by 
introducing the Doppler shift into a BCS Green's function\cite{kubert} 
\begin{equation}
\label{Green}
G({\bf k},\omega_n; {\bf r})=-
{(i\omega_n-{\bf v}_s{\bf k})\tau_0+\Delta_{\bf k}\tau_1+\zeta_{\bf k}\tau_3
\over
(\omega_n+i{\bf v}_s{\bf k})^2+\zeta_{\bf k}^2+\Delta_{\bf k}^2},
\end{equation}
where $\omega_n$ is the fermionic Matsubara frequency, 
$\zeta_{\bf k}$ is the band energy
measured 
with respect to the Fermi level, 
and $\tau_i$ are Pauli matrices. 
Standard many-body techniques can be used to
compute physical quantities $F({\bf r})$ at a fixed position
{\bf r} in real space, and the 
measured quantities are obtained by averaging over the unit cell of the vortex 
lattice according to
\begin{equation}
\label{average}
\langle F\rangle_H={1\over \pi} 
\int_0^1\rho d\rho\int_0^{2\pi}d\beta F(\rho,\beta).
\end{equation}
We first consider the density of states at the Fermi level which is 
easily accessible experimentally via specific heat measurements and
is given by
\end{multicols}
\hrule width 3.45 in \hfill
\begin{eqnarray}
\lb{n0}
N_0(\alpha)&\equiv& -{1\over 2\pi}{\rm Im} \sum_{\bf k} \langle {\rm Tr }G({\bf
k},\omega=0;{\bf r})\rangle_H
\\&=&{1\over 2\pi^2}\int_0^{2\pi} d\phi
	\int_0^{2\pi}d\beta \int_0^1\rho d\rho
\Re {\rm e}\Biggl[{|\sin\beta\sin(\phi-\alpha)|\over
	\sqrt{\sin^2\beta\sin^2(\phi-\alpha)-(\Delta_0/E_H)^2\rho^2 f^2(\phi)}}
	\Biggr].
\end{eqnarray}
The integrals over $\rho$ and $\beta$ can be done analytically, and
the numerical evaluation of Eq.(\ref{n0}) is trivial. 
It has been shown\cite{kubert} that the nodal approximation, which takes
advantage of the fact that the density of states is dominated by the 
contribution of the regions of the Fermi surface
near the gap nodes to replace
${\bf v}_s\cdot{\bf k}$ by the Doppler shift at the nodes,
${\bf v}_s\cdot{\bf k}_n$, provides a remarkably good agreement with
the numerical results for $T,E_H\ll \Delta_0$. Here it yields
\beq
\lb{N0int}
N_0(\alpha)\simeq {E_H \over \D\pi}\sum_{nodes}|\sin(\phi_n-\alpha)|
={2\sqrt2 E_H \over \D\pi}
\max(|\sin\alpha|, |\cos\alpha|).
\seq
\begin{multicols}{2}\noindent
\begin{figure}[h]
\lb{fig:dos}
\epsfxsize=3.27in
\epsfbox{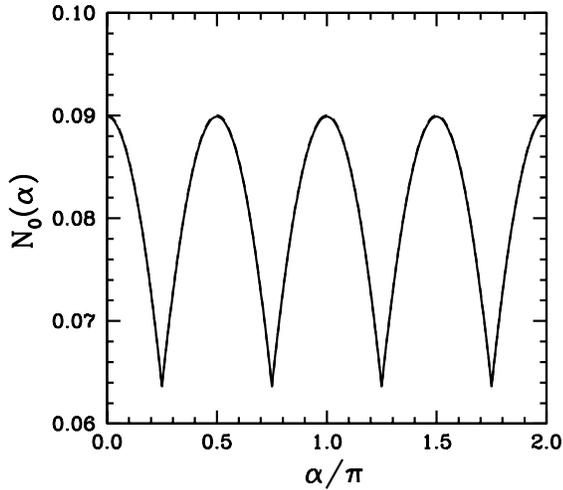}
\narrowtext
\caption{The density of states at the Fermi level, $N_0(\alpha)$, for
$E_H/\D=0.1$. The solid curve is the full numerical evaluation of (5) and 
the dashed, almost indistinguishable, is
the nodal approximation  Eqn.(6).}
\end{figure}
\noindent
This result 
was first obtained by Volovik\cite{volovik2}.
In Fig.1 we show the results of full numerical 
evaluation of the residual density of states from Eq.(\ref{n0}) 
along with the 
results of Eq.(\ref{N0int}).
The density of states 
exhibits 
fourfold oscillations as a function of the angle of the applied field.
There is a broad maximum for the
field applied in the anti-nodal direction, and a sharp minimum for the
field along the node. Both can be understood
if we notice that the Doppler shift
is given by $E_1=E_H|\sin(\pi/4-\alpha)|$ at two of the near-nodal regions, and
by $E_2=E_H|\cos(\pi/4-\alpha)|$ at the other two nodes. When a field is 
applied in the antinodal direction, $E_1=E_2$
and all four nodes contribute
equally to the density of states, as shown in Fig.2a. On the other hand, when 
the field is applied along a nodal direction, 
quasiparticles at that node, which travel
parallel to the field, do not contribute 
to the DOS; the Doppler shift vanishes at 
these points. Moreover, since $E_H\ll\Delta_0$, the gap grows
faster as a function of the angle $\phi$ near the node
than the Doppler shift over most of the unit cell of the vortex lattice, 
and therefore the quasiparticle contribution to the
density of states is suppressed over the whole near-nodal region, see
Fig.2b. For a field not exactly in the nodal direction
there is always a finite region of the momentum space 
where the Doppler shift exceeds the local gap, resulting in a 
contribution to the DOS and sharp minima of $N_0(\alpha)$.

Consequently, 
for a field in the nodal direction two of the nodes do not contribute
to DOS, while the contribution of the other two is a factor of $\sqrt 2$ 
larger than it is for the field along an anti-node. 
\begin{figure}
\lb{fig:FS}
\epsfxsize=3.27in
\epsfbox{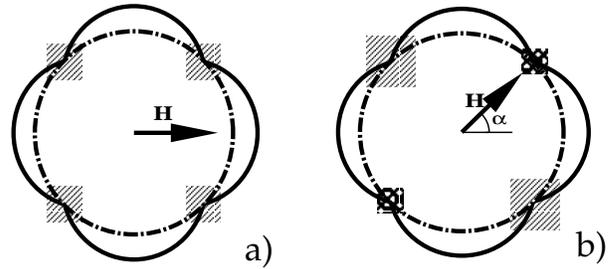}
\caption{\narrowtext
The regions contributing to the density of states for the
anti-nodal (a), and nodal (b) orientation of the magnetic field.
}
\end{figure}\noindent
The density of states is therefore reduced by 
30\%, in agreement with the numerical results.
To check how robust the oscillations are we numerically computed 
$N_0(\alpha)$ in a
 3-dimensional superconductor and found that the results
remains qualitatively unchanged, although the amplitude of the oscillations is
reduced to about 8\%, in agreement with an analytic estimate.
This reduction results from
incomplete suppression of the contribution to the DOS from the
nodal lines aligned with the field since for quasiparticles outside
the equatorial plane the Doppler shift does not vanish. 
We suggest that in realistic materials the amplitude of the
oscillations is somewhere between the two estimates and
within the experimental resolution of the specific heat measurements;
the published results suggest that this amplitude is of order
0.3 mJ/mol K$^2$ at $H=0.005H_{c2||}$\cite{volovik2}.
So far, one reported measurement performed for two orientations of the applied
field in the plane did not find the predicted oscillations\cite{moler}.
However, an estimate shows that
for the samples used in Ref.\cite{moler} the energy scale $E_H$  for $H\sim 8$T
is close to the impurity bandwidth, $\gamma$, which may have resulted in
significant smearing of the fourfold pattern. 
We also note that in an orthorhombic system 
the induced $s$-wave component of the gap 
would shift the position of the DOS minimum away from
the $\pi/4$ direction, and in a heavily twinned
crystal, such as used in Ref.\cite{moler}, this would result in rapid filling
of the minima and significant
suppression of the amplitude of oscillations.

While for a clean sample $N_0(\alpha)\propto\sqrt H$ independently 
of the angular orientation of {\bf H}, the energy dependence of the density
of states
depends crucially on the direction of the field. 
For $\w, E_H \ll \D$,
\beq
\label{dosw}
N(\w,\alpha)\simeq (N_1(\w,\alpha)+N_2(\w,\alpha))/2,
\seq
where \cite{kubert}
\end{multicols}
\widetext
\hrule width 3.5in \hfill
\beq
$$
N_i(\w, \alpha)=\cases{{\w\over\D}\Bigl(1+{1\over 2x^2}\Bigr), &
	if $x=\w/E_i\ge 1$;\cr
	{E_i\over\pi\D x}\Bigl[ (1+2x^2)\arcsin x+ 3x\sqrt{1-x^2}\Bigr],
	& if $x\le 1$\cr}$$,
\seq
\begin{multicols}{2}\noindent
\begin{figure}
\label{fig:wdos}
\epsfxsize=3.27in
\epsfbox{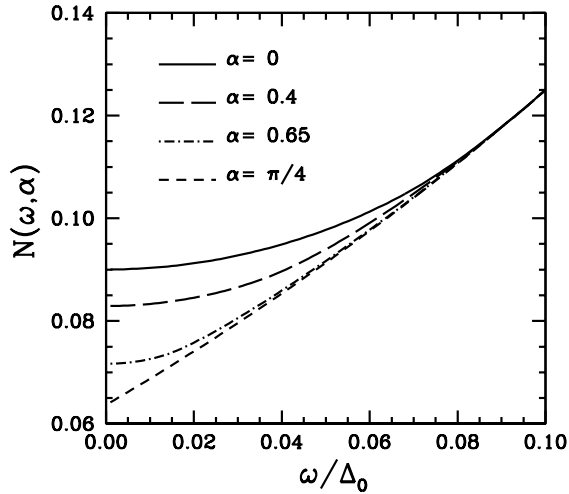}
\caption{\narrowtext
The energy dependent density of states  for different 
directions of
the applied field and $E_H/\Delta_0=0.1$.}
\end{figure}\noindent
for $i=1,2$, and $E_1$ and $E_2$ were defined above.
For the field in the antinodal direction
\beq
\lb{nwan}
N(\w,0)\approx{2\sqrt2 E_H\over \pi\D}
	\Bigl(1+{1\over 3} {\w^2\over E_H^2}\Bigr),
\seq
while for the field along a node
\beq
\lb{nwn}
N(\w, \pi/4)\approx {2 E_H\over \pi\D}+{\w\over 2\D},
\seq
see Fig.3. The frequency dependence of the
density of states follows different power laws
for the field along a node or an anti-node, and in the
former case the linear slope of $N(\w, \pi/4)$
does not depend on the magnetic field. Note that the value of $N(0,0)$
and $N(0,\pi/4)$ differ by a factor of $\sqrt{2}$ as expected  
and that the slope of the linear term in Eq. (10) is just half the value of the
zero field case as only two nodes contribute.

Since the frequency dependence of the density of states determines the
temperature dependence of thermal and transport coefficients, 
our results have profound effects on
the properties of clean 
d-wave superconductors. In particular, in addition to the 
fourfold oscillations
 of the linear-$T$ term in the specific heat as a 
function of the direction of ${\bf H}$, the 
temperature dependence of $C/T\sqrt H$ is linear for the field in a nodal 
direction, and quadratic for {\bf H} away from the node. The nuclear
spin-lattice relaxation time $T_1T$ and
superfluid density will also exhibit
fourfold oscillations and a linear or quadratic 
$T$-dependence depending on the direction of the field. 
However, if the latter quantity is inferred
from the penetration depth measurements, 
nonlocal effects due to a diverging coherence length
 \cite{leggett} in the nodal directions  may be important.

Any orthorhombic distortion
in the system lifts the fourfold degeneracy of the maxima.
One possible way to 
\begin{figure}
\label{fig:ortho}
\epsfxsize=3.27in
\epsfbox{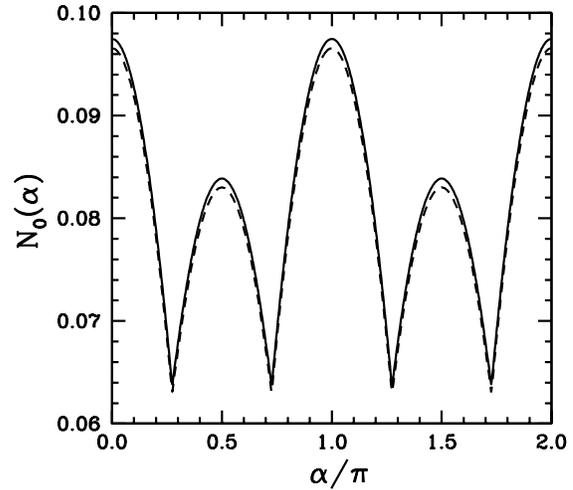}
\caption{\narrowtext
The effect of orthorhombicity included in the gap ($d+0.15s$)
 on $N_0(\alpha)$ for $E_H/\Delta_0=0.1$.
The solid curve is the full numerical evaluation and the dashed line is the
result of the first equality in Eqn. (6).}
\end{figure}\noindent
account for 
such a distortion is to consider a $d+qs$ superconductor, where $q\ll 1$.
Then, even though 
the position of the nodes is shifted insignificantly,
the fourfold pattern 
is replaced by two pairs of peaks with the amplitude
ratio
${|\sin(\alpha-0.5\arccos(-q))|/
		|\sin(\alpha+0.5\arccos(-q))|}$,
which differs significantly from unity  even for relatively
small $q$, as shown in Fig.4.
The anisotropy in the density of states and the thermodynamic 
properties
however remains robust and is a particularly simple
experimental probe of the nodal structure of the order parameter.

We are grateful to K. A. Moler and G. E. Volovik for important
communications.
PJH thanks A. Freimuth and R. Gross for pivotal discussions.
This research 
has been supported in part by NSERC of Canada (EJN and JPC), CIAR (JPC)
and NSF/ AvH Foundation (PJH).
EJN is supported by the 
Cottrell Scholar Program of Research Corporation.

\end{multicols}

\begin{references}

\bibitem{kirtley} J.R. Kirtley {\it et al.}, Nature {\bf 373}, 225 (1995).

\bibitem{annett} J. F. Annett, N. Goldenfeld, and S. R. Renn, in
{\it Physical Properties of High Temperature Superconductors II}, ed. 
by D. M. Ginzberg (World Scientific, Singapore, 1990), and references therein.

\bibitem{volovik} G. E. Volovik, JETP Lett. {\bf 58}, 469 (1993).

\bibitem{kubert} C. K\"ubert and P. J. Hirschfeld, Sol. St. Comm. {\bf 105},
459 (1998).

\bibitem{kubert2} C. K\"ubert and P. J. Hirschfeld,
	Phys. Rev. Lett. {\bf 80}, 4963 (1998).

\bibitem{vekhter} I. Vekhter, J. P. Carbotte, and E. J. Nicol,
	Phys. Rev. B {\bf 59}, 1417 (1999).

\bibitem{barash} Yu. S. Barash, V.P. Mineev, and A.A. Svidzinskii,
        JETP Lett.
        {\bf 65}, 638 (1997).

\bibitem{moler} K. Moler {\it et al.}, Phys. Rev. Lett. {\bf 73}, 2744 (1994);
        Phys. Rev. {\bf B 55}, 3954 (1997).

\bibitem{chiao} M. Chiao {\it et al.}, 
	cond-mat/9810323 (unpublished).

\bibitem{aubin} H. Aubin et al., Phys. Rev. Lett. {\bf 82}, 624 (1999).

\bibitem{sonier} 
	J. E. Sonier {\it et al.}, Phys. Rev. B {\bf 55}, 11789 (1997);

\bibitem{phillips} D. A. Wright {\it et al.}, Phys. Rev. Lett. {\bf 82}, 
	in press.

\bibitem{junod} A. Junod {\it et al.}, Physica C {\bf 282}, 1399 (1997);
B. Revaz {\it et al.}, Phys. Rev. Lett. {\bf 80}, 3364 (1998).

\bibitem{clem1} 
J. R. Clem and M. W. Coffey, Phys. Rev. B {\bf 42}, 6209 (1990).

\bibitem{bulaevskii} L. N. Bulaevskii {\it et al.},
Phys. Rev. B {\bf 50}, 12831 (1994).

\bibitem{clem2} L. N. Bulaevskii and J. R. Clem,  Phys. Rev. B {\bf 44},
		10234 (1991).

\bibitem{bonn} A. Hosseini {\it et al.}, Phys. Rev. Lett. {\bf 81}, 
	1298 (1998).


\bibitem{volovik2} G. E. Volovik, unpublished; in K. A. Moler {\it et al.},
J. Phys. Chem. Solids {\bf 56}, 1899 (1995).

\bibitem{leggett} I. Kosztin and A. Leggett, Phys. Rev. Lett. {\bf 79},
135 (1997).




\end{references}
\end{document}